\documentclass{article}

\usepackage{arxiv}

\usepackage[utf8]{inputenc} % allow utf-8 input
\usepackage[T1]{fontenc}    % use 8-bit T1 fonts
\usepackage{hyperref}       % hyperlinks
\usepackage{url}            % simple URL typesetting
\usepackage{booktabs}       % professional-quality tables
\usepackage{amsfonts}       % blackboard math symbols
\usepackage{nicefrac}       % compact symbols for 1/2, etc.
\usepackage{microtype}      % microtypography
\usepackage{lipsum}		% Can be removed after putting your text content
\usepackage{graphicx}
\usepackage{natbib}
\usepackage{doi}
\usepackage{multirow}

\title{Lightweight Convolution Transformer for Cross-patient Seizure Detection in Multi-channel EEG Signals}

\author{ \href{https://orcid.org/0000-0002-3503-5380}{\includegraphics[scale=0.06]{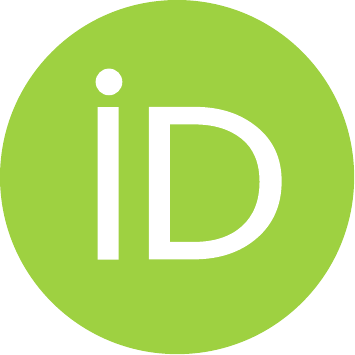}\hspace{1mm}Salim.~Rukhsar}\thanks{Corresponding Author: Salim Rukhsar (rukhsar.1@iitj.ac.in)} \\
	Department of Electrical Engineering\\
	Indian Institute of Technology Jodhpur\\
	India, 342030 \\
	\texttt{rukhsar.1@iitj.ac.in} \\
	\And
	\href{https://orcid.org/0000-0000-0000-0000}{\includegraphics[scale=0.06]{orcid.pdf}\hspace{1mm}Anil K. Tiwari} \\
	Department of Electrical Engineering\\
	Indian Institute of Technology Jodhpur\\
	India, 342030 \\
	\texttt{akt@iitj.ac.in} \\
}

\date{}

\renewcommand{\headeright}{}
\renewcommand{\undertitle}{}
\renewcommand{\shorttitle}{LIGHTWEIGHT CONVOLUTION TRANSFORMER FOR SEIZURE DETECTION}

\hypersetup{
pdftitle={Lightweight Convolution Transformer for Cross-patient Seizure Detection in Multi-channel EEG Signals},
pdfsubject={q-bio.NC, q-bio.QM},
pdfauthor={Salim Rukhsar, Anil K. Tiwari},
pdfkeywords={Epilepsy, Seizure detection, Vision Transformer, Convolutional Neural Network, Electroencephalogram},
}

\begin{document}
\maketitle

\begin{abstract}
\emph{\textbf{Background:}} Epilepsy is a neurological illness affecting the brain that makes people more  likely to experience frequent, spontaneous seizures. There has to be an accurate automated method for measuring seizure frequency and severity in order to assess the efficacy of pharmacological therapy for epilepsy. The drug quantities are often derived from patient reports which may cause significant issues owing to inadequate or inaccurate descriptions of seizures and their frequencies.\\
\emph{\textbf{Methods and materials:}} This study proposes a novel deep learning  architecture based lightweight convolution transformer (LCT). The transformer is able to learn spatial and temporal correlated information simultaneously from the multi-channel electroencephalogram (EEG) signal to detect seizures at smaller segment lengths. In the proposed model, the lack of  translation equivariance and localization of ViT is reduced using convolution tokenization, and rich information from the transformer encoder is extracted by sequence pooling instead of the learnable class token.\\
\emph{\textbf{Results:}} Extensive experimental results demonstrate that the proposed model of cross-patient learning can effectively detect seizures from the raw EEG signals. The accuracy and F1-score of seizure detection in the cross-patient case on the CHB-MIT dataset are shown to be 96.31$\%$ and 96.32$\%$, respectively, at 0.5 sec segment length. In addition, the performance metrics show that the inclusion of inductive biases and attention-based pooling in the model enhances the performance and reduces the number of transformer encoder layers, which significantly reduces the computational complexity. In this research work, we provided a novel approach to enhance efficiency and simplify the architecture for multi-channel automated seizure detection.
\end{abstract}

% keywords can be removed

\section{Introduction}
Epilepsy is a non-infectious, recurring, chronic neurological condition marked by abnormal and sudden alterations in brain electrical activity \citep{WHO2016}. Recent statistics suggest that around fifty million persons are suffering all over the world from epilepsy, resulting in one of the most frequent neurological illnesses globally \citep{RUKHSAR2019320}. Drug-resistant epilepsy is diagnosed when two or more effective trials of anti-epileptic medication fail to control the seizure of patients \citep{LOPEZGONZALEZ2015439}. The neurological examination is performed on certain patients to see if they are good candidates for non-pharmacological alternative treatments such as neurosurgery, the ketogenic diet, or responsive neurostimulation. In spite of these options, there is still a sizable population of people who continue to experience recurrent seizures. There is a strong correlation between the severity of seizures and the severity of their effects on these patients' quality of life and mortality rates. This is something that can affect both women and men of any age, class, race, or location \citep{HU2020103919}. Nonetheless, it has a major impact on the lives of both young adults in their twenties (the prime of their productive life) and the elderly (due to higher risk). The risk of early death for those with epilepsy is three times higher than that of the general population \citep{LACHAKE2021507,WHO2016,LOPEZGONZALEZ2015439}.

It is estimated that seventy-eighty percent of seizures can be managed and twenty five percent of seizures can be prevented with the help of anticonvulsant medication \citep{YUAN201799}. Protecting the head from harm and reducing the incidence of epilepsy in children caused by birth injuries are the most effective ways to prevent post-traumatic epilepsy. The anomalies in electroencephalogram (EEG) signal pattern and a well-documented aetiology of the seizure are the two most dependable indications of seizure recurrence \citep{Siddiqui201799,RUKHSAR2019320}. A reliable automated approach for the estimation of seizure frequency and severity is essential in assessing the efficacy of drug treatment in epilepsy. These quantities are often derived from patient reports which may result in significant issues owing to inadequate or inaccurate descriptions of seizures and their frequencies. Seizure detection needs a comprehensive inspection and analysis of huge quantities of EEG data by neurologists, but the quality of the analysis is heavily reliant on their expertise and experience of the neurologist. Examination using sight alone takes a lot of time and might lead to erroneous conclusions because of human subjectivity \citep{PENG2021104338,LOPEZGONZALEZ2015439}. Consequently, an accurate seizure detection system for epilepsy is necessary for clinical and scientific applications.

There has been a lot of focus and significant advancement in recent years on the development of new technology for automated seizure event identification. Several methods based on linear and non-linear properties have been developed since J. Gotman proposed the first automated seizure detection system in 1982 \citep{GOTMAN1982530}. Many supervised and unsupervised "machine learning" algorithms have been developed for seizure classification from EEG signals of interictal states. The ability of support vector machine (SVM) techniques to decrease structural risk has led to their greater popularity in recent years for the classification of manually produced features \citep{GOSHVARPOUR2022105240,FU201415,RUKHSAR2023104833}. Using ANNs, \citep{YAVUZ2018201} were able to accurately distinguish between seizure and non-seizure cases. During the past few years, much of the focus has been on deep learning (DL) approaches due to their effective results in solving such problems \citep{9139257}. The end consequence of using DL approaches to find an incorporate patterns from processed or raw data in a system that is highly reliable but at a significantly higher computational cost. In addition, because EEG data are time-series signals, i.e., progressively monitored across time, temporal and spatial correlation can be found for seizure identification. Fortunately, Deep Learning has shown to be a method with substantial promise for evaluating such sort of data, as it permits the learning of long-term properties and represents the state-of-the-art (SOTA) for many works of time series classification \citep{Hassandeep}.

Convolutional neural networks (CNN) \citep{Jemalcross},\citep{ACHARYA2018270} to analyse visual images and recurrent neural networks (RNN) \citep{HUSSEIN201925} for sequential data are the widely used deep learning methods. These techniques will automatically identify and extract the essential features from unprocessed data to produce the required results. In an RNN, the hidden state at each time step depends on the previous time step's hidden state and the current input. As a result, RNNs are limited in their ability to process long sequences, as the hidden state must retain information from previous time steps, and this information can be lost or diluted over time. \citep{vaswani2017attention} developed a transformer architecture to perform sequential data processing in parallel, which is inherently impossible with the RNN and solves the long training time problems of CNN. The transformer uses a self-attention mechanism to compute the importance of each element in the input sequence. This allows the model to attend to all elements in the sequence simultaneously, regardless of their position. As a result, the transformer can process long sequences of data more efficiently and effectively without losing important information or experiencing the vanishing gradient problem. Additionally, the transformer's architecture allows for more parallelism than an RNN. In an RNN, the computation at each time step depends on the previous time step, making it difficult to parallelize the computations across different time steps. In contrast, the transformer's self-attention mechanism allows for parallel computation across all elements in the sequence, making it possible to process the sequence much more quickly. Currently, the transformer has become a de-facto standard for natural language processing (NLP) tasks but its application in computer vision is limited.
Because of the limited application, \citep{dosovitskiy2021image} proposed a vision transformer to classify images into their categories. Despite the fact that multi-channel EEG signals contain sequential data, the transformer design has not yet been applied to their processing for seizure. Regarding EEG signal processing, \cite{krishna2020eeg} presented a transformer-based approach for automated speech recognition based on EEG characteristics. Using a transformer model to predict the response to several anti-epileptic drugs in people with a new diagnosis of epilepsy yielded promising results \citep{Choong2020.11.10.20229385}. \citep{Sun9858598} used a traditional vision transformer for seizure detection in long-term iEEG, in which classification tokens are utilised to categorise the data. These latest developments provide a broader scope of applications, which in turn helps to defend the use of transformers for time series data.

Although some encouraging results have been documented employing such approaches, there are still certain challenges that need to be overcome. To begin with, most static models do not take into account the dynamic correlations between each channels and timestamp data, thus fail to identify spatial and temporal correlations simultaneously. Second, the use of traditional deep learning features with a simplistically designed architecture may lose information and thus cannot  guarantee reliable results. This is especially true when dealing with datasets that are not evenly distributed. Lastly, developing a cross-patient detector is challenging since seizure patterns in EEG data can vary greatly between patients and even over time for the same patient.

To address the aforementioned issues, we present a novel seizure detection method that utilises the spatial and temporal information of multi-channel EEG data from both intra and inter-patient groups. The proposed framework is described in Figure \ref{my_figure3}. We extracted ictal and interictal data of equal length from each patient and concatenated them respectively to make a cross-patient dataset. The prepared dataset is normalized and segmented with different segment lengths at 25$\%$ overlapping. Here are the contributions in a nutshell:
\begin{itemize}
  \item The proposed framework is capable of learning spatial and temporal correlated information simultaneously from multi-channel EEG signals for cross-patient seizure classification.
  \item We developed a novel lightweight convolution transformer (LCT) by introducing inductive biases, which the transformer lacks compared to CNN and attention-based pooling. The developed architecture used convolution tokenizer instead of patches and attention-based pooling instead of classification token. The LCT is able to escape from the big data paradigm of transformers and is successful in the effective classification of smaller data. The results show that the proposed method outperforms the state-of-the-art methods with less learnable parameters.
\end{itemize}

The remaining portions of this work are divided as follows: Section 2 is dedicated to the EEG dataset, pre-processing, and transformer architecture. The proposed architecture of the lightweight convolution transformer (LCT) is elaborated deeply in section 3. The experimental results are presented in section 4. An in-depth discussion and comparison with other methods are presented in section 5. Finally, section 6 outlines concluding remarks.

\section{Methodology}
In this research work, an automated seizure detection system using a novel lightweight convolution transformer (LCT) is proposed. The EEG data is initially processed and then segmented in a multi-channel form that makes the window in a 2D form similar to images. LCT consists of two novel architectures compared to ViT, these two are sequence pooling and convolutional tokenizer. The LCT is evaluated on the segmented ictal and interictal portion of CHB-MIT scalp multi-channel EEG dataset.

\subsection{Data description}
Long-term scalp EEG recordings were obtained at the Children's Hospital Boston and made available to the public through \href{https://physionet.org/content/chbmit/1.0.0/}{https://physionet.org} as a CHB-MIT scalp EEG dataset. This dataset was utilised to assess the efficacy of the study described by \citep{SHOEB2004483}. Twenty-three  paediatric patients with intractable seizure disorders provided the data for this recording. The 21 electrodes used to record 18-23 channels of electrical brain activity are labelled according to the international 10-20 electrode placement standard. The signal has been band-pass filtered between 0 to 128 Hz and sampled at 256 Hz. Chb-24 data is excluded from the experiment due to the unavailability of suitable interictal periods.
Patients diagnosed with epilepsy often experience seizures that last for a considerably shorter amount of time in comparison to their interictal period. A common problem in detecting seizures in EEG data is an imbalance between the number of ictal and interictal periods \citep{YUAN201799}. Classifiers tend to favour the class that has the most data segments due to a bias introduced during the training process of classification models. In order to create a balanced dataset, it was decided that there should be the same amount of interictal as ictal segments. For this reason, only eighteen consistent channels C3-P3, F7-T7, P7-O1, FP1-F7, FP1-F3, P3-O1, FP2-F8, T7-P7, F3-C3, T8-P8, F4-C4, F8-T8, FP2-F4, CZ-PZ, P8-O2, C4-P4, FZ-CZ, and P4-O2 are considered in this
work \citep{RUKHSAR2019320}.

\subsection{Data preprocessing and segmentation}
Before the training phase, the EEG dataset is pre-processed by applying z-score normalisation to the stacked datasets of each individual patient for normal and seizure class. This normalization is performed on all channels simultaneously to ensure that data has mean $\mu_{X}$ and standard deviation $\sigma_{X}$, according to (\ref{eqn1}).
\begin{equation}
     \hat{X} = \frac{X-\mu_{X}}{\sigma_{X}}
    \label{eqn1}
\end{equation}
 A single EEG segment is represented as a matrix of dimension $(N \times L)$ where $N$ denotes the number of channels, and $L$ is the sequence length = $256 \times $ segment duration. Segment length of [0.0625, 0.125, 0.25, 0.50, 0.75, 1.0, 1.5, and 2.0] sec with $25\%$ overlapping segment is used for evaluating the proposed architecture. As an example, a 1 sec segment is represented as an $18\times 256$ dimensional matrix, with N = 18 chosen for the reasons explained earlier. The segments of the EEG dataset are then formed by putting all the ictal and interictal of all 23 patients in one matrix whose dimension is $(2K\times N\times L)$, where $K$ represents the number of the interictal or ictal segments.
 \subsection{Transformer Architecture}
Transformers are structures designed to handle sequential signals and are the cutting edge for natural language processing (NLP) jobs. Despite the fact that the current research has explored its applications from computer vision \citep{dosovitskiy2021image} to drug response prediction \citep{ott-etal-2018-scaling}, the same is yet not much explored with multi-channel time series signals, such as scalp EEG. When comparing NLP with EEG processing, it can be seen that both include sequential data, and an event from a short-term EEG segment of a time series signal can be understood in the context of the  word vectors in NLP. Similarly, the representation of a sentence in matrix form can be similar to a multichannel EEG segment, where the single channel data can be considered as a one-hot vector.

\begin{figure}
    \centering
    \centerline{\includegraphics[width=17cm, height=8cm]{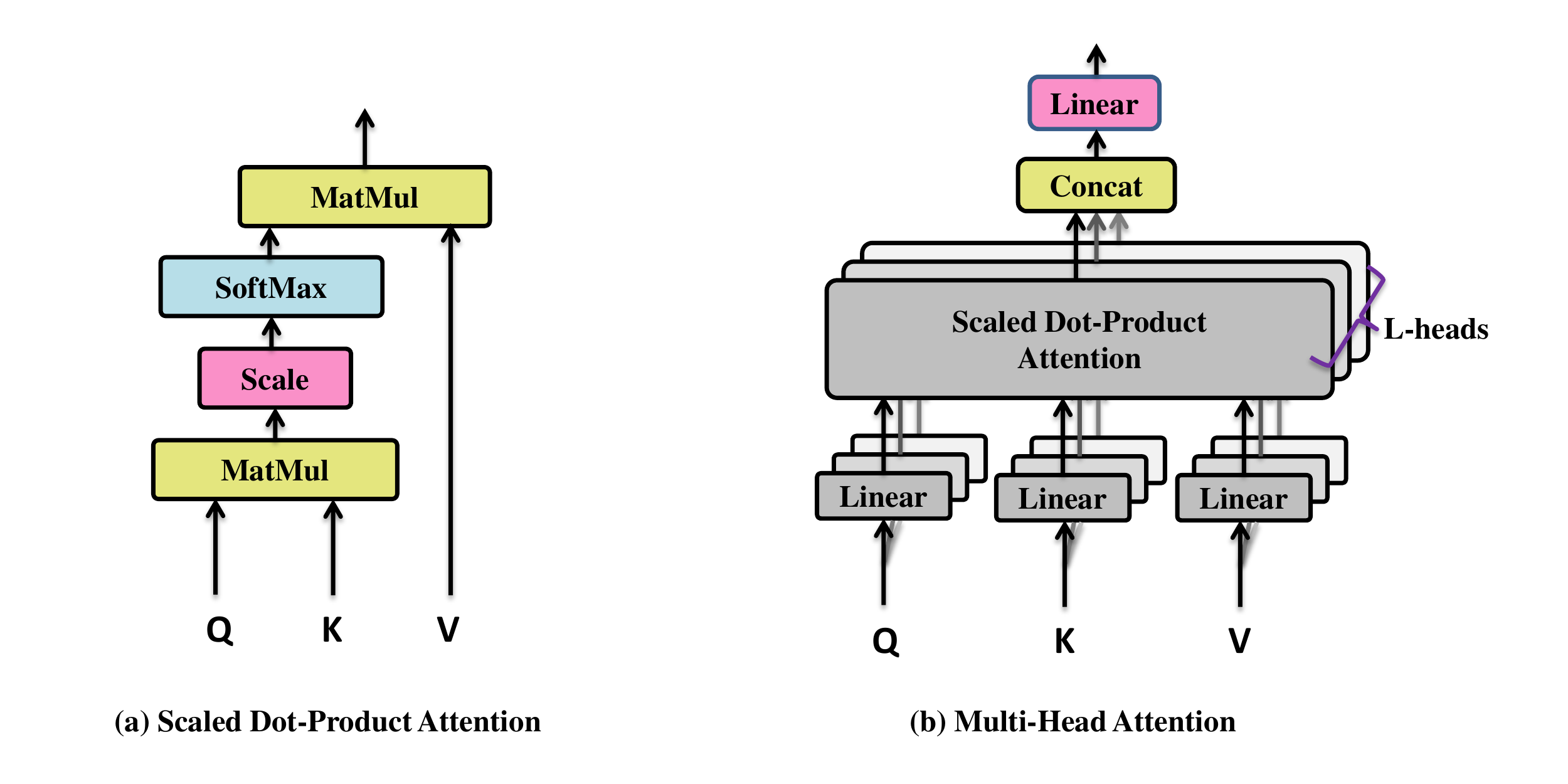}}
    \caption{The detailed structure of transformer module: (a) Scaled dot-product attention, (b) Multi-head attention}
    \label{my_figure1}
\end{figure}

The transformer is free from any convolution or recurrence architecture and it simply relies on attention mechanism. This architecture uses an encoder and decoder with an attention mechanism to concentrate on the regions that are of utmost importance for a particular input and thus devote more computing resources to these regions. \citep{vaswani2017attention} used the Scaled Dot-Product attention as depicted in Fig.\ref{my_figure1}(a). The attention mechanism multiplied the input vector by three weight matrices $[W^{Q},W^{K},W^{V}]$ to produce the queries vector, keys vector, and values vector designated as Q, K, and V, respectively, is described in (\ref{eqn2}). The dimensions of the three vectors are denoted by $d_{q}, d_{k}$, and $d_{v}$ respectively. The self-attention (SA) of data sequence $\hat{X}$ can be calculated by taking the softmax of scaled dot-product of the queries (Q) and keys (K) by $\sqrt{d_{k}}$ and multiplying the values vector (V), as expressed in (\ref{eqn3}).
\begin{equation}
       [Q,K,V] = \hat{X}[W^{Q},W^{K},W^{V}]
    \label{eqn2}  
\end{equation}
\begin{equation}
     SA(\hat{X}) = softmax \left(\frac{QK^{T}}{\sqrt{d_{k}}}\right)V
    \label{eqn3}
\end{equation}
Where the projections are parameter matrices $W^{Q} \in \mathcal{R}^{d_{model}\times d_{k}}$, $W^{K} \in \mathcal{R}^{d_{model}\times d_{k}}$, and $W^{V} \in \mathcal{R}^{d_{model}\times d_{v}}$.
The “Scaled Dot Product Attention" layers that comprise multi-head attention allowed the model to jointly focus on data from several representation sub-spaces at various places \citep{vaswani2017attention}. This “Multi-Head Attention" is shown in Fig \ref{my_figure1}(b) and can be expressed as:
\begin{equation}
       MultiHead(\hat{X}) = Concat(SA_{1}(\hat{X}),\dots, SA_{h}(\hat{X}))W^{o}
    \label{eqn4}  
\end{equation}
 Where the projections are parameter metric $W^{o} \in \mathcal{R}^{h\times d_{v}\times d_{model}}$. In this encoder architecture, the vector dimensions $d_{k}= d_{v}= d_{model}/h =64$ for h=2.

\subsection{Vision Transformer}
\begin{figure}
    \centering
    \centerline{\includegraphics[width=17cm, height=9cm]{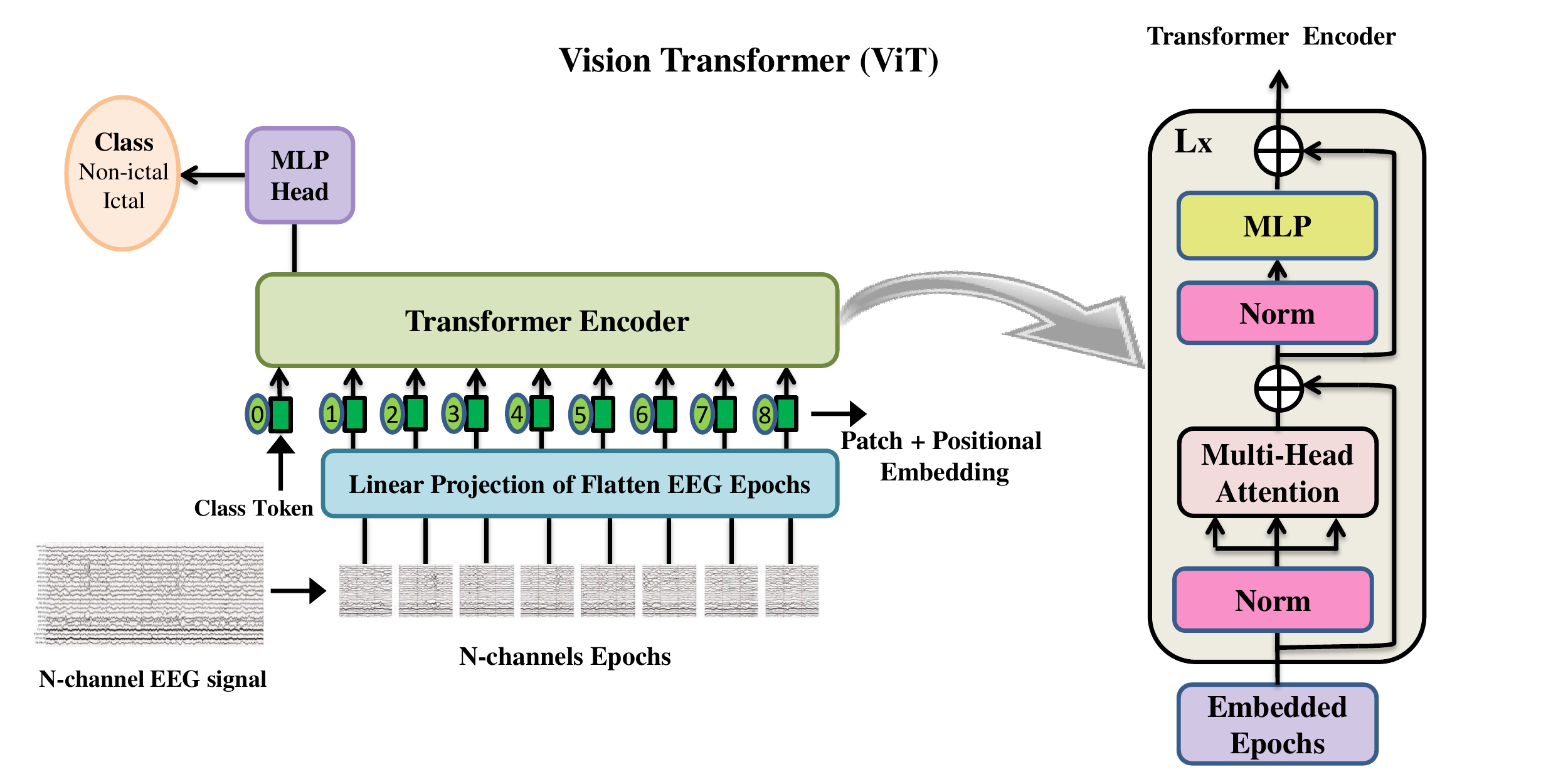}}
    \caption{In the transformer encoder, norm stands for normalization, and MLP as multilayer perceptron}
    \label{my_figure2}
\end{figure}

Vision transformer (ViT) is a transformer architecture that has been modified to take images as input. ViT can therefore handle 2D images as input rather than 1D sequential data. It begins by generating 2D patches from the input, and then sends these patches and their associated positional embeddings onto a transformer encoder (see Fig. \ref{my_figure2}). A class token was used by \citep{dosovitskiy2021image} for classification by adding it to the linear projection of the flattened patches. This input  stack of the sequence is processed by the transformer encoder, and then the class token is used by the multilayer perceptron (MLP) layer for the classification of the corresponding class labels. The Multi-Head Attention mechanism in ViT stacks data from each head of the model and concatenates it into a single output. 

\section{Lightweight Transformer}
We adopt the original transformer \citep{dosovitskiy2021image} and  vision transformer \citep{vaswani2017attention} models to create the proposed model architecture. The transformer encoder consists of a normalization layer, multi-head self-attention layer, dropout layer, and MLP layer. To mitigate the vanishing gradient, residual connections are present in the encoder. The transformers do not generalise well when trained on an insufficient quantity of data because they lack several of the inductive biases present in CNNs, such as translation equivariance and localization \citep{Sun9858598} \citep{hassani2022escaping}. Therefore, the inclusion of CNN layers in the transformer may lead to a lightweight transformer. Extraction of relevant information from the encoder output by using attention-based pooling may enhance the performance and eliminate the use of class token from ViT.

\begin{figure}
    \centering
    \centerline{\includegraphics[width=18cm, height=12cm]{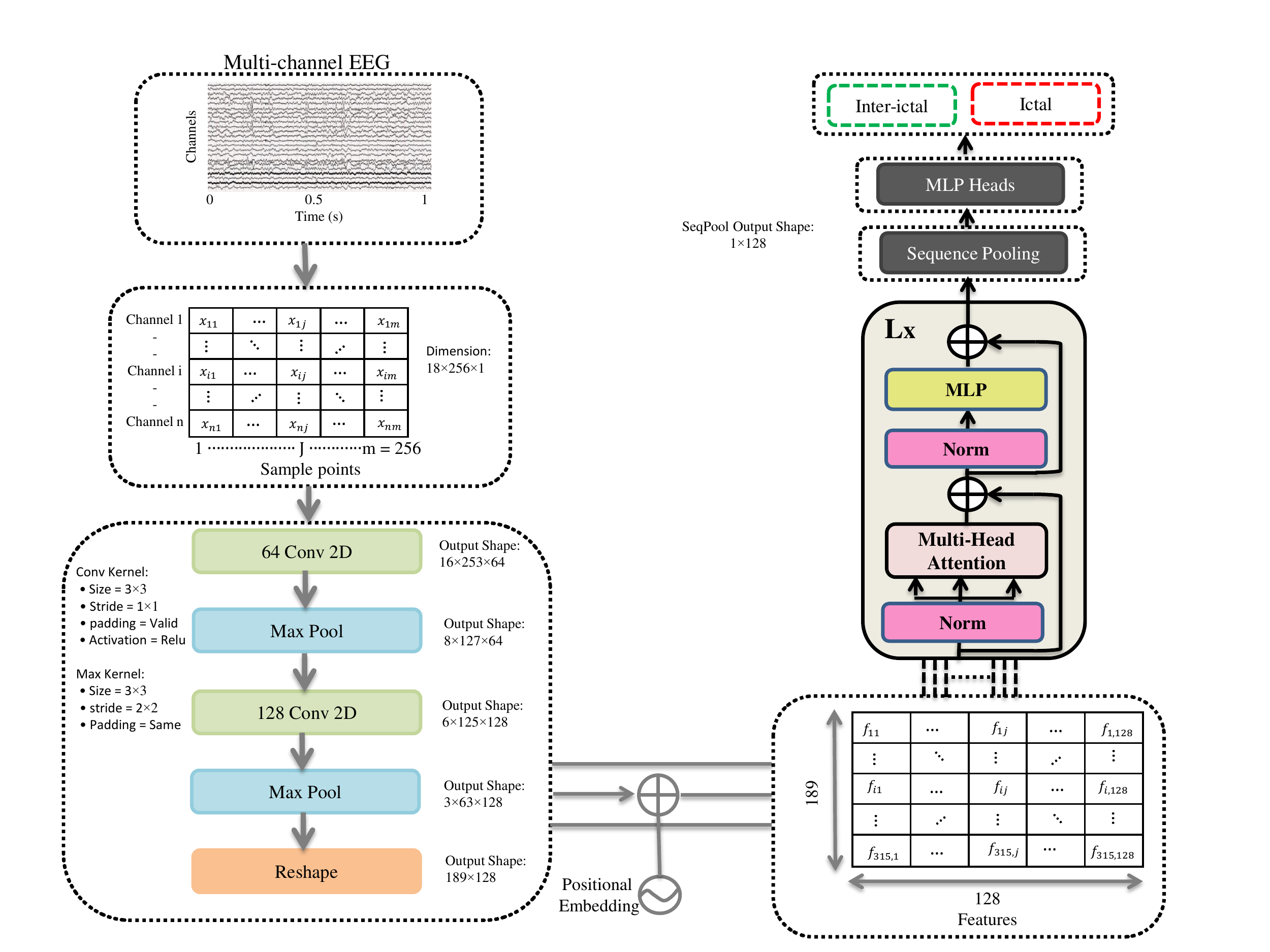}}
    \caption{Proposed lightweight convolution transformer (LCT) architecture and flow diagram}
    \label{my_figure3}
\end{figure}

\subsection{Sequence Pooling}
Learnable class token has been used in ViT and other common transformer-based classifiers to map the sequential output to a singular class index, and later feeding it to the classifier. Global average pooling is another frequent approach that has been demonstrated to be superior in various circumstances. We present sequence pooling (SeqPool), an attention-based technique that pools over the output token sequence. Motivation for the use of SeqPool is that the output sequence contains relevant information across different channels of the EEG signals. Therefore, preserving this information can enhance performance with no additional requirements of the class token. Additionally, by forwarding one less token, this adjustment somewhat reduces computation. The output sequence mapped in this operation using the transformation $T : \mathcal{R}^{b\times n\times d} \rightarrow \mathcal{R}^{b\times d}$, can be expressed:

\begin{equation}
     Y_{L} = f(\hat{X}) \in \mathcal{R}^{b\times n\times d}
    \label{eqn5}
\end{equation}

where $a_{L}$ is the result of an $L$-layer transformer encoder $f$, $b$ is the size of the batch, $n$ is the number of sequences, and $d$ is the overall dimension of the embedding. $a_{L}$ is fed to a linear layer $g(a_{L}) \in \mathcal{R}^{d\times 1}$, and softmax activation is applied to it, as expressed in (\ref{eqn6})
\begin{equation}
     \hat{Y}  = softmax(g(Y_{L})^{T}) \in \mathcal{R}^{b\times 1\times n}
    \label{eqn6}
\end{equation}
 This provides a significant weighting for each input sequence, which can be expressed as in the case of self-attention as expressed in (\ref{eqn7}).
\begin{equation}
     S  = \hat{Y}Y_{L} = softmax(g(Y_{L})^{T})\times Y_{L} \in \mathcal{R}^{b\times 1\times d}
    \label{eqn7}
\end{equation}

By reshaping the output to $S\in \mathcal{R}^{b\times d}$ and the reshaped output is then sent to MLP for classification. The SeqPool is able to make the model to weigh the embedding of latent space produced by transformer encoder and correlate data across the different input sequence. This may be understood as paying attention to the sequential data, where important weights are assigned to the sequence of data after the encoder has processed it. The addition of the SeqPool block to the ViT eliminates the need of a class token. Consequently, a computationally efficient architecture is created that can be denoted as a lightweight vision transformer (LVT).

\subsection{Convolution Tokenizer}\label{section 3.2}
We altered the LVT by replacing its patch embedding approach with a straightforward convolution block in order to incorporate an inductive bias into the model. This block follows a conventional design, which consists of two single convolution layers, ReLU activation, and a max pool layer. For a given 2D signal or feature map $\hat{X} \in \mathcal{R}^{H\times W\times C}$:
\begin{equation}
     \hat{X}_{0} = maxpool\left(ReLU(conv2d(\hat{X}))\right)
    \label{eqn8}
\end{equation}
Where $H$, $W$, and $C$ represent the height, width, and channel respectively in case of image. In this work, $H$ represents the number of EEG Channels, $W$ is the length of the segment, and $C =1$. Here the 2D convolution (Conv2d) operation has $d$ filters, the same number as the embedding dimension of the transformer architecture. Furthermore, the max pool and convolution operations may overlap, which might boost the performance by adding inductive biases. This enables the proposed model to preserve local spatial information. Furthermore, by incorporating this convolutional block, the model have more flexibility than models like ViT. Because of the convolution tokenizer, the LCT is no longer restricted to input resolutions that are precisely divided by the pre-set patch size. As the convolutions will be more effective, and hence resulting into richer tokens for the transformer, we employed to embed the multi-channel EEG signal into a latent representation. The down sampling of the input feature map results in fewer tokens, which significantly reduces computations since self-attention has a quadratic time and space complexity with respect to the token size. The proposed model architecture and its flow diagram are shown in Fig. \ref{my_figure3}

The lightweight convolution transformer (LCT) is created by the novel convolutional tokenizer along with sequence pooling. We used the notation of LCT variants with the number of transformer encoder layers and heads of MHA. For instance, LCT-$3/2$ has three encoder layers, and two heads of MHA.

\section{Experimental Results}
The ictal and interictal data are extracted from all patients and concatenated in single seizure and interictal class except chb-24 due to insufficient interictal data in it. To ensure consistency, channels are chosen that are common among all patients. There are only eighteen such channels, namely P7-O1,  FP1-F7, FP1-F3, C3-P3, F7-T7, P3-O1, FP2-F8, T7-P7, F3-C3, F8-T8, FP2-F4, T8-P8, F4-C4, CZ-PZ,  P8-O2, P4-O2,C4-P4, and FZ-CZ, considered in this work. The processed multi-channel EEG dataset is segmented using different time durations $[ 0.0625, 0.125, 0.25, 0.50, 0.75, 1.0, 1.5, 2.0]$ secs with $25\%$ overlapping. For a single segment of length 1 sec of data, the segment dimension can be $18\times 256$ with $64$ overlapping data points. The segmented dataset, in this case, is reshaped into (Number of segments, 18, 256, 1). Furthermore, the segmented dataset is split into $75\%$ and $25\%$ for training and testing respectively. The $10\%$ of the raining data is furthermore assigned for validating the model, and the remaining is used for fine-tuning of weights and biases. 

To achieve optimal performance and prevent over-fitting, the training process is repeated multiple times, upto 300 epochs, and is stopped when the specified stopping criteria is met. Further, the optimization of the model during the training process is accomplished by adjusting the hyperparameters, as given in Table \ref{tab1}. One optimization technique that is used is step decay, which gradually reduces the learning rate at set intervals (every 25 epochs) by a specific factor (0.1) starting from the initial value (0.001) using the Adam optimizer. Lastly, the model is updated after processing a batch of 300 input segments.

\begin{table}[!ht]
\resizebox{\textwidth}{!}{
\centering
\begin{tabular}{c c c c c c c } \hline
 Input Size & Optimizer & Learning rate & Learning rate schedule & Learning rate drop & Batch size & Epochs  \\ \hline 
$(18\times 256\times 1)$  & Adam & $1e^{-3}$ & Step decay & $1e^{-1}$ & 300 & 300   \\ \hline       
\end{tabular}}
  \caption{Hyperparameters used for optimization during training the models for one second segment length}
  \label{tab1}
   \vspace{-3pt}
\end{table}

To show that vision transformers can perform better than convolutional neural networks, even when the data sets are small, we compare the ViT with and without convolution tokenizer. Initially, ViT is modified by adding sequential pooling (SeqPool) to the transformer encoder instead of a class token. After that, the resulting vision transformer is further modified with the convolution tokenizer instead of patches. Addition of SeqPool in ViT results in a lightweight vision transformer (LVT), and further modification of the convolution tokenizer results in a lightweight convolution transformer (LCT).\\
\textbf{Vision Transformer (ViT):} In this work, we implemented the vision transformer proposed by \citep{dosovitskiy2021image}  for seizure detection using the multi-channel time series data. The number of samples in the segment is taken $252$ samples $\approx 1$ sec with $0.25$ sec overlapping. The dimension of the constructed EEG segmented data is (Number of segments $\times 18\times 252\times 1)$. One segment is treated as 2D images, and 14 patches of dimension $18\times 18$ are extracted. The extracted patches are flattened ($18\times18 = 324$) and projected into $529$ to make the dimension consistent (Number of segments $\times 14 \times 529)$. Furthermore, a learnable class token is concatenated with the reshaped patches. \\
\textbf{Lightweight Vision Transformer (LVT):} The performance of ViT with very few number of encoder layers is relatively not good even though it has a lot of parameters. To make the ViT able to escape from the big data paradigm, we need to boost the performance of ViT by making appropriate modifications to its architecture. The first modification introduced in the ViT is the sequence pooling (SeqPool). The SeqPool processing can be thought of as paying attention to the sequential data, where importance to the weights are being assigned across the sequence of data of different channels, after the transformer encoder has processed it. The SeqPool extracts relevant information and eliminates the use of the  class token.\\
\textbf{Lightweight Convolution Transformer (LCT):} Since the ViT and LVT used the patches of 2D EEG signal, there is an inherent issue of information loss between the edges of the patches. This issue is resolved by using a max pooling with overlapping convolution layers as described in section \ref{section 3.2}. The proposed model should have some relational bias, as a result, the transformer becomes capable of learning from the information that is more thoroughly integrated. In this convolution tokenization, we used two blocks of overlapping convolution followed by a maxpool with 32 and 128 filters in the respective blocks. The 2D conv layer is constructed using kernel size = $(3\times 3)$, stride = $(1\times 1)$, padding = valid, and activation = Relu with max pool parameters of kernel size = $(3\times 3)$, stride = $(2\times 2)$, and padding = same. This layer transforms one segment EEG data of dimension ($18 \times 256\times 1)$ to ($189 \times 128)$.

\begin{figure}
    \centering
    \centerline{\includegraphics[width=16cm, height=8cm]{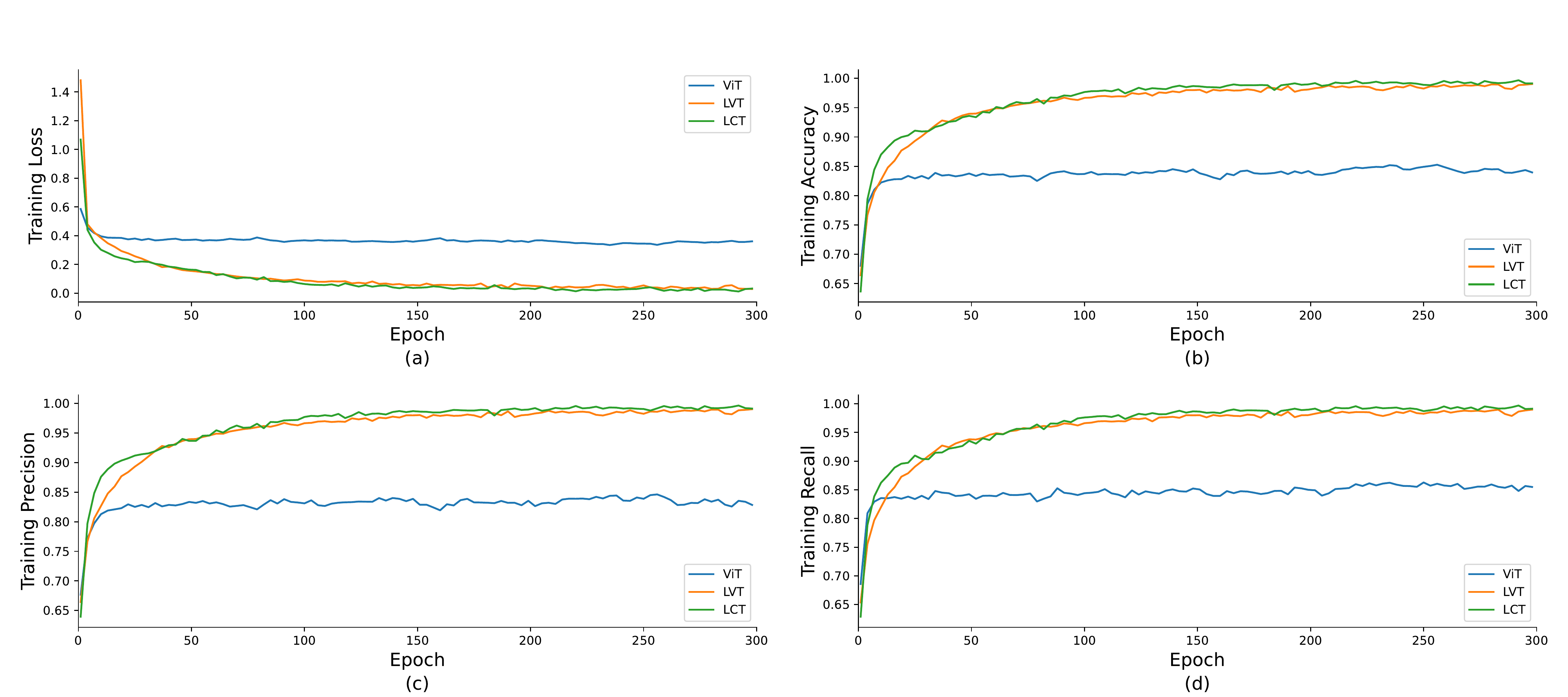}}
    \caption{Performance parameters of the proposed models with ViT}
    \label{my_figure4}
\end{figure}

\begin{figure}
    \centering
    \centerline{\includegraphics[width=16cm, height=8cm]{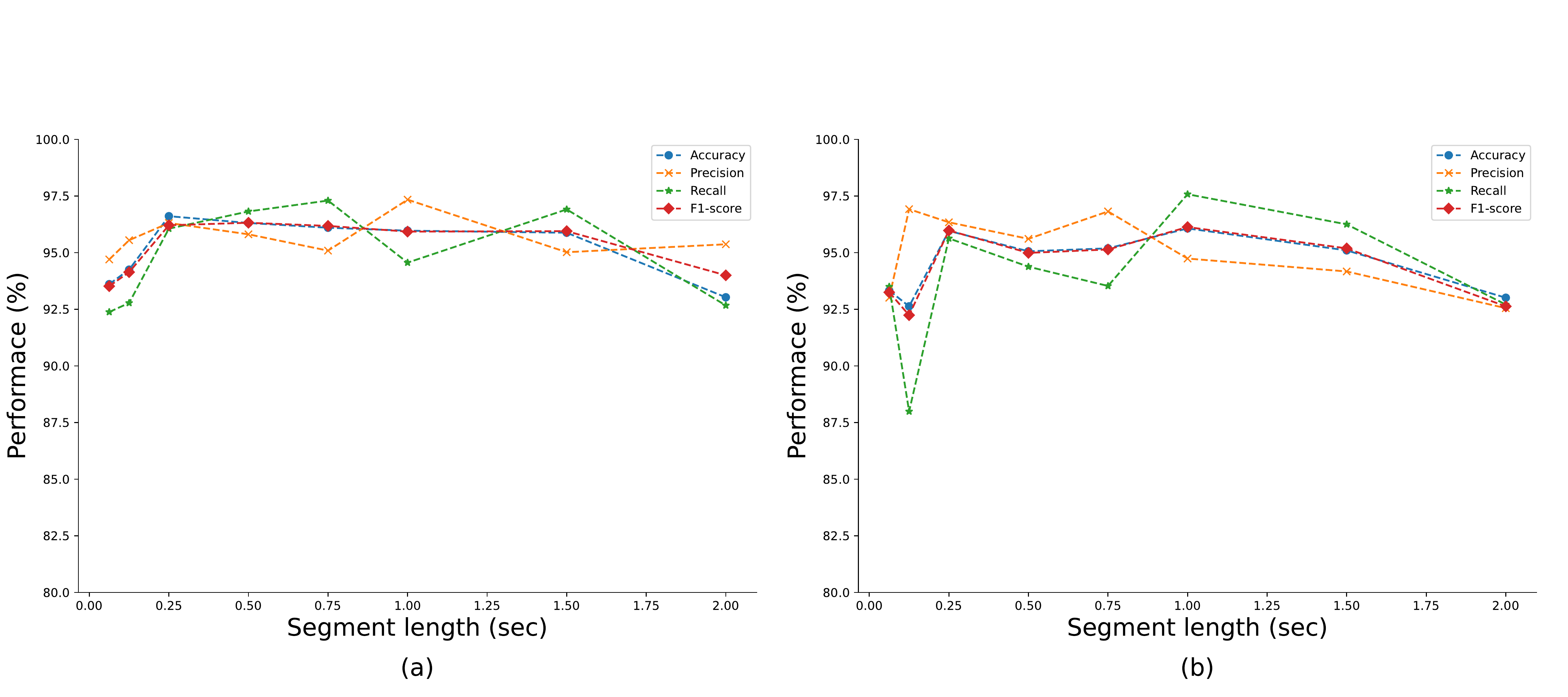}}
    \caption{Performance indicators at different segment lengths by the proposed model (a) LCT 1/2, and (b) LCT 2/2.}
     \vspace{-1pt}
    \label{my_figure5}
\end{figure}

 To evaluate the performance, the findings are reported using standard performance indicators so that we may assess how well the proposed model works. The performance indicators used are accuracy, precision, recall, and F1-score. We have evaluated the proposed models on the publicly available CHB-MIT dataset. Table \ref{tab2} summarizes the performance of the VIT and two modified version of it with different encoder layers and multi-head attention. From this table, it can be observed how well SeqPool and then convolution tokenizer enhance the performance of the transformer for seizure detection. A detailed performance analysis as a function of different segment lengths is presented in Fig. \ref{my_figure5}. From this, it can be observed, that a consistent performance 
 is found for the segment length 0.5 sec for different variants of LCT as given in Table \ref{tab3}. From the performance for the two variants, i.e. LCT 1/2 and LCT 2/2, the best segment length for the detection found is 0.5 sec because the performance at this length is consistent among all variants. The best performance and faster responses of LCT are at smaller segment lengths  due to the quadratic computational complexity of MHA. This is also a desired characteristic of a model for seizure detection to provide a response at a faster rate.
 
Based on these observations, a comparative analysis of the performance of the proposed model with state-of-the-art methods on the same dataset is shown in Table \ref{table4}. As far as we are aware, very few studies have been conducted using the CHB-MIT dataset for cross-patient seizure detection. \citep{8217737} used a context-learning based approach for fusing multi-features. The fused features are classified using SVM method and reported 95.71$\%$ accuracy. On the same dataset, another cross-patient detection work was reported by Jemal et al. \citep{Jemalcross}. In this work, a CNN based on separable depthwise convolution is proposed for automatic cross-subject seizure detection. The proposed LCT model outperformed the SOTA methods with less computational complexity and higher detection accuracy. There are many seizure detection works on the same CHB-MIT dataset reported, but these works are for patient-specific cases only. \citep{8395430} developed an improved transductive transfer learning Takagi-Sugeno-Kang fuzzy system (ETTL-TSK-FS) and obtained a classification accuracy of 94.00$\%$. The channel-embedding spectral-temporal squeeze-and-excitation network (CE-stSENet) was introduced by \citep{Li8995501}, reported detection accuracy of 95.96$\%$. Recently, \citep{ZHAO2023104441} proposed a model based on CNN and transformer to detect seizures. They reported an accuracy of 98.76$\%$, by using CNN for local feature extraction, and the transformer for global features. The comparative performance of the proposed model with ViT is shown in Fig. \ref{my_figure4}. In conclusion, our system provided a higher sensitivity, recall, and F1-score for cross-patient seizure detection on the CHB-MIT dataset.
    
\begin{table}[!ht]
\resizebox{\textwidth}{!}{
\centering
\begin{tabular}{l | c ||c c c c } \hline
 Model & Parameters & Accuracy (\%) & Precision (\%) & Recall (\%) & F1-score (\%)  \\ \hline 
\multirow{5}{*}{ViT} &{1/2} & 86.19 & 87.28 & 86.65 & 0.869  \\  
        &{2/2}  & 89.07 & 90.52 & 89.56 &  0.900  \\ 
        &{3/2}  & 83.73 & 85.43 & 81.79 & 0.836   \\
        &{4/2}  & 73.88 & 74.67 & 73.21 & 0.739   \\
        &{4/3}  & 71.69 & 73.17 & 77.06 & 0.751   \\ \hline
\multirow{5}{*}{LVT} & {1/2}  & 94.93 & 94.64 & 94.45& 0.945  \\ 
        &{2/2}  & 94.47 & 93.72 & 95.28 & 0.945  \\ 
        &{3/2}  & 90.41 & 91.22 & 89.49 & 0.903   \\
        &{4/2}  & 84.34 & 85.01 & 83.42 & 0.842   \\
        &{4/3}  & 81.12 & 81.50 & 80.58 & 0.810   \\ \hline
\multirow{5}{*}{LCT} & \textbf{1/2}  & \textbf{95.91} & \textbf{94.87} & \textbf{95.23} & \textbf{0.951} \\ 
        &{2/2}  & 94.73 & 95.11 & 94.12 & 0.946  \\ 
        &{3/2}  & 94.15 & 94.43 & 93.13 & 0.938   \\
        &{4/2}  & 93.79 & 92.62 & 93.29 & 0.929   \\
        &{4/3}  & 93.70 & 92.52 & 92.42 & 0.925   \\ \hline
        
\end{tabular}}
  \caption{The performance of each model with a different number of encoder layers and heads for 300 epochs and 1 $sec$ segment length.}
  \label{tab2}
   \vspace{-1pt}
\end{table}

\begin{table}[!ht]
\resizebox{\textwidth}{!}{
\centering
\begin{tabular}{l | c ||c | c | c } \hline
 Model & Performance parameters & W= 0.5 sec/OL=0.125 sec &  W= 1 sec/OL=0.25 sec &  W= 2 sec/OL= 0.5 sec \\ \hline 
\multirow{4}{*}{\textbf{LCT 1/2}} & {Accuracy}  & \textbf{96.31} & 95.97 & 93.03   \\ 
    &{Precision}  & 95.81 & 97.34 & 95.37   \\ 
    &{Recall}  & 96.82 & 94.56 & 92.67   \\
    &{F1-score}  & \textbf{0.963} & 0.959 & 0.940  \\ \hline
\multirow{4}{*}{\textbf{LCT 2/2}} & {Accuracy}  & 95.06 & 96.07 & 93.01    \\ 
        &{Precision}  & 95.61 & 94.74 & 92.55   \\ 
        &{Recall}  & 94.38 & 97.58 & 92.71   \\ 
        &{F1-score}  & 0.949 & 0.961 & 0.926   \\\hline
\multirow{3}{*}{\textbf{LCT 3/2}} & {Accuracy}  & 95.16 & 95.77 & 92.26    \\ 
        &{Precision}  & 95.41 & 95.25 & 92.49   \\ 
        &{Recall}  & 94.81 & 96.38 & 93.16   \\ 
        &{F1-score}  & 0.951 & 0.958 & 0.928   \\\hline
\multirow{3}{*}{\textbf{LCT 4/2}} & {Accuracy}  & 94.61 & 93.69 & 92.15    \\ 
        &{Precision}  & 93.57 & 93.71 & 91.12   \\ 
        &{Recall}  & 92.55 & 93.05 & 90.81  \\ 
        &{F1-score}  & 0.930 & 0.934 & 0.909   \\\hline
\end{tabular}}
\footnotesize{ W denotes the length of the segment and OL is the length of overlapping .}
  \caption{The performance of LCT variants at different window lengths with 25$\%$ overlapping.}
  \label{tab3}
   %\vspace{-1pt}
\end{table}

\section{Discussion}
In the proposed method, a novel lightweight convolution transformer is proposed based on a vision transformer. The proposed work shows that a ViT can be more effective with a smaller dataset by combining the attention-based pooling and the ability of feature extraction by convolutional tokenizer. Since the use of transformers in natural language processing has been found to be effective, we explored it for EEG time series signals. The ability of transformer architecture to extract relevant information available in different EEG channels at different time instants, is explored for seizure detection. Although transformer models perform well on their own, in this work, it is found that fusion models that include  convolutional layers and sequence pooling further boost their performances. This is likely attributable to the fact that CNN adds inductive biases in the model and is, therefore quite effective in feature extraction. Including positional embedding (PE) prior to the transformer processing has been shown to considerably enhance the accuracy of the model. The results on the CHB-MIT dataset by the proposed model, show significant better performance as compared to competitive methods for cross-patients cases.

Several researchers have trained seizure classification algorithms using EEG data from individual patients \citep{salim10040195}. Patient-specific detection often involves combining data from many channels to gather additional samples and splitting it into multiple trials of the same subject to a training and testing set. While the patient-specific approach yields better accuracy, the trained model is not well suited on other patients. In this work, a generalized model for seizure detection is developed by working on cross-patient data for the training and validation, on twenty-three patients. Adaptability and resilience are key features of this extended model, making it suitable for usage with diverse characteristics of patients and more general-purpose EEG dataset-based tasks.

Prior transformer-based approaches for seizure detection work mostly focused on patient-specific cases, and not on cross-patient seizure identification, with smaller learnable parameters of the transformer. In this work, we have introduced inductive biases as present in CNNs, such as translation equivariance and localization, by using a convolution tokenizer instead of patches. An attention-based pooling technique is used to extract relevant information from the output token sequence that contains relevant information across different channels of the EEG signal. Therefore the inclusion of SeqPool and convolution tokenizer makes the ViT capable of learning the relevant information quickly and, therefore, effectively classifies the ictal and interictal with a single encoder layer.

Initially, the raw dataset is normalized by applying z-score normalization. To prepare the data into 2D form, multi-channel EEG data at different segment lengths are prepared with 25$\%$ overlapping. We presented ViT and modified forms of it, namely LVT, and LCT, for seizure detection. The performance parameters of all the variants with different encoder layers and heads are given in Table \ref{tab2}. From these results, it can be seen that the inclusion of SeqPool and convolution tokenizer significantly improved the classification result. The best result by ViT model at two encoder layers and two heads (ViT 2/2) is $89.07\%$. Addition of SeqPool before the classification MLP highly improved the classification results with only one encoder layer. The results show that the token-based ViT looses information, and hence more encoder layers are required. Also, this shows the superiority of attention-based pooling that leads to lightweight transformers due to the reduction in the number of encoder layers. Furthermore, improvement is seen by the inclusion of inductive biases of CNN layer as a convolution tokenizer instead of patches. Moreover, the proposed model gains flexibility by employing this convolutional tokenizer block as opposed to models like ViT, which are limited to an input segment length that is exactly divided by the preset patch size. The addition of SeqPool and convolution tokenizer transformed the ViT into a lightweight convolution transformer (LCT). The details of performance of LCT 1/2 and LCT 2/2 as a function of different segment lengths is shown in fig. (\ref{my_figure5}). At 0.5-sec segment length, a performance of 96.31$\%$ accuracy and 0.9632 F1-score is achieved by LCT 1/2. The performance of different variants of LCT is shown in Table \ref{tab3}. It is observed that LCT performs better at smaller segment lengths, consequently, the computation speed is faster due to the quadratic computational complexity of MHA $\mathcal{O}(P^2)$, where $P$ is the the patch size.This is advantageous in the case of seizure detection because we are able to detect seizures at smaller intervals, i.e., 0.5 sec. While classification with small segment data is challenging, LCT at a smaller input segment improved the detection accuracy of the seizure detection system. This demonstrates the robustness and reliability of LCT for other applications based on multi-channel EEG signals. 
 \begin{table}[!ht]
\resizebox{\textwidth}{!}{
\centering
\begin{tabular}{c c c c c c c} \hline\hline
\textbf{Authors} & \textbf{Year} & \textbf{Method} & \textbf{Classification case} & \textbf{Accuracy}& \textbf{Precision} & \textbf{Recall}\\ \hline

\citep{Li8995501} & {2020} & {CE-stSENet} & patient-specific & 95.96  & 96.05 & 92.41 \\\hline

\citep{8395430} & {2018} & {TTL-FSs} & patient-specific & 94.00  & 93.20 & 91.90 \\\hline

\citep{ZHAO2023104441} & {2023} & {CNN+Transformer} & patient-specific & 98.76  & 97.60 & 97.70 \\\hline\hline

\citep{8217737} & {2017} & {WT-CtxFusionEEG} & cross-patient & 95.71  & 96.08 & 98.65 \\\hline

 \citep{Jemalcross}& {2021} & {CNN} & cross-patient & 91.82  & - & 91.93 \\\hline

\textbf{Ours} & {2023} & {LCT} & cross-patient & 96.31  & 95.81 & 96.82 \\\hline
\end{tabular}}
\caption{Analysis of the proposed work in comparison to other SOTA methods on the same CHB-MIT dataset.}
\label{table4}
\end{table}

\section{Conclusion}
In this work, a lightweight convolution transformer (LCT) architecture that automatically learns spatial and temporal information from multi-channel scalp EEG signals simultaneously is proposed. The raw multi-channel scalp EEG data is used as the input for the model, and it does not require any extra feature engineering, pre-set patching, and learnable classification token. The proposed model included inductive biases and attention-based pooling to extract relevant information and eliminates the information loss at the edge of patches. It is also reduced to a single encoder layer, and working better on smaller segment lengths, consequently reducing the computational cost. The LCT model is capable of learning the correlated information between channels and different time instants simultaneously. Tests results of the cross-patient cases on the CHB-MIT dataset at different segment lengths and LCT variants validate the reliability and robustness of the model. It is hoped that the findings of this study will encourage the further development of technologies for real time seizures detection system. This LCT model has the potential to meet the strong need for a simpler architecture for seizure prediction. The proposed method is generic enough that can be adopted for other tasks based on EEG data-base as well.

\section*{Declaration of competing interest}
The authors declare that they have no conﬂicts of interest.

\bibliographystyle{unsrtnat}
\bibliography{references}  %%% Uncomment this line and comment out the ``thebibliography'' section below to use the 

\begin{thebibliography}{31}
\providecommand{\natexlab}[1]{#1}
\providecommand{\url}[1]{\texttt{#1}}
\expandafter\ifx\csname urlstyle\endcsname\relax
  \providecommand{\doi}[1]{doi: #1}\else
  \providecommand{\doi}{doi: \begingroup \urlstyle{rm}\Url}\fi

\bibitem[WHO(2023)]{WHO2016}
World health oraganization (2016) epilepsy [cited may 6, 2023].
\newblock \emph{World Health Oraganization}, 2023.
\newblock URL
  \url{https://www.who.int/en/news-room/fact-sheets/detail/epilepsy}.

\bibitem[Rukhsar et~al.(2019)Rukhsar, Khan, Farooq, Sarfraz, and
  Khan]{RUKHSAR2019320}
S.~Rukhsar, Y.U. Khan, O.~Farooq, M.~Sarfraz, and A.T. Khan.
\newblock Patient-specific epileptic seizure prediction in long-term scalp eeg
  signal using multivariate statistical process control.
\newblock \emph{IRBM}, 40\penalty0 (6):\penalty0 320--331, 2019.
\newblock ISSN 1959-0318.
\newblock \doi{https://doi.org/10.1016/j.irbm.2019.08.004}.
\newblock URL
  \url{https://www.sciencedirect.com/science/article/pii/S1959031818302719}.

\bibitem[{López González} et~al.(2015){López González}, {Rodríguez
  Osorio}, {Gil-Nagel Rein}, {Carreño Martínez}, {Serratosa Fernández},
  {Villanueva Haba}, {Donaire Pedraza}, and {Mercadé
  Cerdá}]{LOPEZGONZALEZ2015439}
F.J. {López González}, X.~{Rodríguez Osorio}, A.~{Gil-Nagel Rein},
  M.~{Carreño Martínez}, J.~{Serratosa Fernández}, V.~{Villanueva Haba},
  A.J. {Donaire Pedraza}, and J.M. {Mercadé Cerdá}.
\newblock Drug-resistant epilepsy: Definition and treatment alternatives.
\newblock \emph{Neurología (English Edition)}, 30\penalty0 (7):\penalty0
  439--446, 2015.
\newblock ISSN 2173-5808.
\newblock \doi{https://doi.org/10.1016/j.nrleng.2014.04.002}.
\newblock URL
  \url{https://www.sciencedirect.com/science/article/pii/S2173580815001091}.

\bibitem[Hu et~al.(2020)Hu, Yuan, Xu, Leng, Yuan, and Yuan]{HU2020103919}
Xinmei Hu, Shasha Yuan, Fangzhou Xu, Yan Leng, Kejiang Yuan, and Qi~Yuan.
\newblock Scalp eeg classification using deep bi-lstm network for seizure
  detection.
\newblock \emph{Computers in Biology and Medicine}, 124:\penalty0 103919, 2020.
\newblock ISSN 0010-4825.
\newblock \doi{https://doi.org/10.1016/j.compbiomed.2020.103919}.
\newblock URL
  \url{https://www.sciencedirect.com/science/article/pii/S0010482520302614}.

\bibitem[Lachake et~al.(2021)Lachake, Desai, and Udani]{LACHAKE2021507}
Aishwarya~V. Lachake, Neelu Desai, and Vrajesh Udani.
\newblock ‘to reveal or to conceal’- disclosure strategies in parents of
  children with epilepsy in india.
\newblock \emph{Seizure}, 91:\penalty0 507--512, 2021.
\newblock ISSN 1059-1311.
\newblock \doi{https://doi.org/10.1016/j.seizure.2021.07.026}.
\newblock URL
  \url{https://www.sciencedirect.com/science/article/pii/S1059131121002570}.

\bibitem[Yuan et~al.(2017{\natexlab{a}})Yuan, Zhou, Zhang, Zhang, Xu, Leng,
  Wei, and Chen]{YUAN201799}
Qi~Yuan, Weidong Zhou, Liren Zhang, Fan Zhang, Fangzhou Xu, Yan Leng, Dongmei
  Wei, and Meina Chen.
\newblock Epileptic seizure detection based on imbalanced classification and
  wavelet packet transform.
\newblock \emph{Seizure}, 50:\penalty0 99--108, 2017{\natexlab{a}}.
\newblock ISSN 1059-1311.
\newblock \doi{https://doi.org/10.1016/j.seizure.2017.05.018}.
\newblock URL
  \url{https://www.sciencedirect.com/science/article/pii/S1059131117303795}.

\bibitem[Siddiqui et~al.(2020)Siddiqui, Morales-Menendez, Huang, and
  Hussain]{Siddiqui201799}
Mohammad~Khubeb Siddiqui, Ruben Morales-Menendez, Xiaodi Huang, and Nasir
  Hussain.
\newblock A review of epileptic seizure detection using machine learning
  classifiers.
\newblock \emph{Brain Inf}, 7:\penalty0 1--18, 2020.
\newblock \doi{https://doi.org/10.1186/s40708-020-00105-1}.
\newblock URL
  \url{https://braininformatics.springeropen.com/articles/10.1186/s40708-020-00105-1}.

\bibitem[Peng et~al.(2021)Peng, Lei, Zheng, Zhao, Wu, Sun, and
  Hu]{PENG2021104338}
Hong Peng, Chang Lei, Shuzhen Zheng, Chengjian Zhao, Chunyun Wu, Jieqiong Sun,
  and Bin Hu.
\newblock Automatic epileptic seizure detection via stein kernel-based sparse
  representation.
\newblock \emph{Computers in Biology and Medicine}, 132:\penalty0 104338, 2021.
\newblock ISSN 0010-4825.
\newblock \doi{https://doi.org/10.1016/j.compbiomed.2021.104338}.
\newblock URL
  \url{https://www.sciencedirect.com/science/article/pii/S0010482521001323}.

\bibitem[Gotman(1982)]{GOTMAN1982530}
J~Gotman.
\newblock Automatic recognition of epileptic seizures in the eeg.
\newblock \emph{Electroencephalography and Clinical Neurophysiology},
  54\penalty0 (5):\penalty0 530--540, 1982.
\newblock ISSN 0013-4694.
\newblock \doi{https://doi.org/10.1016/0013-4694(82)90038-4}.
\newblock URL
  \url{https://www.sciencedirect.com/science/article/pii/0013469482900384}.

\bibitem[Goshvarpour and Goshvarpour(2022)]{GOSHVARPOUR2022105240}
Atefeh Goshvarpour and Ateke Goshvarpour.
\newblock A novel 2-piece rose spiral curve model: Application in epileptic eeg
  classification.
\newblock \emph{Computers in Biology and Medicine}, 142:\penalty0 105240, 2022.
\newblock ISSN 0010-4825.
\newblock \doi{https://doi.org/10.1016/j.compbiomed.2022.105240}.
\newblock URL
  \url{https://www.sciencedirect.com/science/article/pii/S0010482522000324}.

\bibitem[Fu et~al.(2014)Fu, Qu, Chai, and Dong]{FU201415}
Kai Fu, Jianfeng Qu, Yi~Chai, and Yong Dong.
\newblock Classification of seizure based on the time-frequency image of eeg
  signals using hht and svm.
\newblock \emph{Biomedical Signal Processing and Control}, 13:\penalty0 15--22,
  2014.
\newblock ISSN 1746-8094.
\newblock \doi{https://doi.org/10.1016/j.bspc.2014.03.007}.
\newblock URL
  \url{https://www.sciencedirect.com/science/article/pii/S1746809414000457}.

\bibitem[Rukhsar and Tiwari(2023)]{RUKHSAR2023104833}
Salim Rukhsar and Anil~Kumar Tiwari.
\newblock Barnes–hut approximation based accelerating t-sne for seizure
  detection.
\newblock \emph{Biomedical Signal Processing and Control}, 84:\penalty0 104833,
  2023.
\newblock ISSN 1746-8094.
\newblock \doi{https://doi.org/10.1016/j.bspc.2023.104833}.
\newblock URL
  \url{https://www.sciencedirect.com/science/article/pii/S1746809423002665}.

\bibitem[Yavuz et~al.(2018)Yavuz, Kasapbaşı, Eyüpoğlu, and
  Yazıcı]{YAVUZ2018201}
Erdem Yavuz, Mustafa~Cem Kasapbaşı, Can Eyüpoğlu, and Rıfat Yazıcı.
\newblock An epileptic seizure detection system based on cepstral analysis and
  generalized regression neural network.
\newblock \emph{Biocybernetics and Biomedical Engineering}, 38\penalty0
  (2):\penalty0 201--216, 2018.
\newblock ISSN 0208-5216.
\newblock \doi{https://doi.org/10.1016/j.bbe.2018.01.002}.
\newblock URL
  \url{https://www.sciencedirect.com/science/article/pii/S0208521617303716}.

\bibitem[Rasheed et~al.(2021)Rasheed, Qayyum, Qadir, Sivathamboo, Kwan,
  Kuhlmann, O’Brien, and Razi]{9139257}
Khansa Rasheed, Adnan Qayyum, Junaid Qadir, Shobi Sivathamboo, Patrick Kwan,
  Levin Kuhlmann, Terence O’Brien, and Adeel Razi.
\newblock Machine learning for predicting epileptic seizures using eeg signals:
  A review.
\newblock \emph{IEEE Reviews in Biomedical Engineering}, 14:\penalty0 139--155,
  2021.
\newblock \doi{10.1109/RBME.2020.3008792}.

\bibitem[Fawaz et~al.(2019)Fawaz, Forestier, Weber, Idoumghar, and
  Muller]{Hassandeep}
Hassan~Ismail Fawaz, Germain Forestier, Jonathan Weber, Lhassane Idoumghar, and
  Pierre-Alain Muller.
\newblock Deep learning for time series classification: a review.
\newblock \emph{Data Min Knowl Disc}, 33:\penalty0 917–963, 2019.
\newblock \doi{https://doi.org/10.1007/s10618-019-00619-1}.
\newblock URL
  \url{https://link.springer.com/article/10.1007/s10618-019-00619-1}.

\bibitem[Jemal et~al.(2021)Jemal, Mitiche, Abou-Abbas, Henni, and
  Mezghani]{Jemalcross}
Imene Jemal, Amar Mitiche, Lina Abou-Abbas, Khadidja Henni, and Neila Mezghani.
\newblock An effective deep neural network architecture for cross-subject
  epileptic seizure detection in eeg data.
\newblock In \emph{Frontiers in Artificial Intelligence and Applications},
  volume 345, pages 54 -- 62, 2021.
\newblock \doi{https://doi.org/10.3233/FAIA210389}.
\newblock URL \url{https://ebooks.iospress.nl/volumearticle/58755}.

\bibitem[Acharya et~al.(2018)Acharya, Oh, Hagiwara, Tan, and
  Adeli]{ACHARYA2018270}
U.~Rajendra Acharya, Shu~Lih Oh, Yuki Hagiwara, Jen~Hong Tan, and Hojjat Adeli.
\newblock Deep convolutional neural network for the automated detection and
  diagnosis of seizure using eeg signals.
\newblock \emph{Computers in Biology and Medicine}, 100:\penalty0 270--278,
  2018.
\newblock ISSN 0010-4825.
\newblock \doi{https://doi.org/10.1016/j.compbiomed.2017.09.017}.
\newblock URL
  \url{https://www.sciencedirect.com/science/article/pii/S0010482517303153}.

\bibitem[Hussein et~al.(2019)Hussein, Palangi, Ward, and Wang]{HUSSEIN201925}
Ramy Hussein, Hamid Palangi, Rabab~K. Ward, and Z.~Jane Wang.
\newblock Optimized deep neural network architecture for robust detection of
  epileptic seizures using eeg signals.
\newblock \emph{Clinical Neurophysiology}, 130\penalty0 (1):\penalty0 25--37,
  2019.
\newblock ISSN 1388-2457.
\newblock \doi{https://doi.org/10.1016/j.clinph.2018.10.010}.
\newblock URL
  \url{https://www.sciencedirect.com/science/article/pii/S1388245718313464}.

\bibitem[Vaswani et~al.(2017)Vaswani, Shazeer, Parmar, Uszkoreit, Jones, Gomez,
  Kaiser, and Polosukhin]{vaswani2017attention}
Ashish Vaswani, Noam Shazeer, Niki Parmar, Jakob Uszkoreit, Llion Jones,
  Aidan~N. Gomez, Lukasz Kaiser, and Illia Polosukhin.
\newblock Attention is all you need, 2017.

\bibitem[Dosovitskiy et~al.(2021)Dosovitskiy, Beyer, Kolesnikov, Weissenborn,
  Zhai, Unterthiner, Dehghani, Minderer, Heigold, Gelly, Uszkoreit, and
  Houlsby]{dosovitskiy2021image}
Alexey Dosovitskiy, Lucas Beyer, Alexander Kolesnikov, Dirk Weissenborn,
  Xiaohua Zhai, Thomas Unterthiner, Mostafa Dehghani, Matthias Minderer, Georg
  Heigold, Sylvain Gelly, Jakob Uszkoreit, and Neil Houlsby.
\newblock An image is worth 16x16 words: Transformers for image recognition at
  scale, 2021.

\bibitem[Krishna et~al.(2020)Krishna, Tran, Carnahan, and
  Tewfik]{krishna2020eeg}
Gautam Krishna, Co~Tran, Mason Carnahan, and Ahmed~H Tewfik.
\newblock Eeg based continuous speech recognition using transformers, 2020.

\bibitem[Choong et~al.(2020)Choong, Hakeem, Chen, Brodie, Lawn, Drummond, Kwan,
  and Ge]{Choong2020.11.10.20229385}
Jiun Choong, Haris Hakeem, Zhibin Chen, Martin Brodie, Nicholas Lawn, Tom
  Drummond, Patrick Kwan, and Zongyuan Ge.
\newblock Application of transformers for predicting epilepsy treatment
  response.
\newblock \emph{medRxiv}, 2020.
\newblock \doi{10.1101/2020.11.10.20229385}.
\newblock URL
  \url{https://www.medrxiv.org/content/early/2020/11/13/2020.11.10.20229385}.

\bibitem[Sun et~al.(2022)Sun, Jin, Si, Zhang, Cao, Wang, Yin, and
  Ming]{Sun9858598}
Yulin Sun, Weipeng Jin, Xiaopeng Si, Xingjian Zhang, Jiale Cao, Le~Wang, Shaoya
  Yin, and Dong Ming.
\newblock Continuous seizure detection based on transformer and long-term ieeg.
\newblock \emph{IEEE Journal of Biomedical and Health Informatics}, 26\penalty0
  (11):\penalty0 5418--5427, 2022.
\newblock \doi{10.1109/JBHI.2022.3199206}.

\bibitem[Shoeb et~al.(2004)Shoeb, Edwards, Connolly, Bourgeois, {Ted Treves},
  and Guttag]{SHOEB2004483}
Ali Shoeb, Herman Edwards, Jack Connolly, Blaise Bourgeois, S.~{Ted Treves},
  and John Guttag.
\newblock Patient-specific seizure onset detection.
\newblock \emph{Epilepsy \& Behavior}, 5\penalty0 (4):\penalty0 483--498, 2004.
\newblock ISSN 1525-5050.
\newblock \doi{https://doi.org/10.1016/j.yebeh.2004.05.005}.
\newblock URL
  \url{https://www.sciencedirect.com/science/article/pii/S1525505004001593}.

\bibitem[Ott et~al.(2018)Ott, Edunov, Grangier, and
  Auli]{ott-etal-2018-scaling}
Myle Ott, Sergey Edunov, David Grangier, and Michael Auli.
\newblock Scaling neural machine translation.
\newblock In \emph{Proceedings of the Third Conference on Machine Translation:
  Research Papers}, pages 1--9, Brussels, Belgium, October 2018. Association
  for Computational Linguistics.
\newblock \doi{10.18653/v1/W18-6301}.
\newblock URL \url{https://aclanthology.org/W18-6301}.

\bibitem[Hassani et~al.(2022)Hassani, Walton, Shah, Abuduweili, Li, and
  Shi]{hassani2022escaping}
Ali Hassani, Steven Walton, Nikhil Shah, Abulikemu Abuduweili, Jiachen Li, and
  Humphrey Shi.
\newblock Escaping the big data paradigm with compact transformers, 2022.

\bibitem[Yuan et~al.(2017{\natexlab{b}})Yuan, Xun, Jia, and Zhang]{8217737}
Ye~Yuan, Guangxu Xun, Kebin Jia, and Aidong Zhang.
\newblock A novel wavelet-based model for eeg epileptic seizure detection using
  multi-context learning.
\newblock In \emph{2017 IEEE International Conference on Bioinformatics and
  Biomedicine (BIBM)}, pages 694--699, 2017{\natexlab{b}}.
\newblock \doi{10.1109/BIBM.2017.8217737}.

\bibitem[Deng et~al.(2018)Deng, Xu, Xie, Choi, and Wang]{8395430}
Zhaohong Deng, Peng Xu, Lixiao Xie, Kup-Sze Choi, and Shitong Wang.
\newblock Transductive joint-knowledge-transfer tsk fs for recognition of
  epileptic eeg signals.
\newblock \emph{IEEE Transactions on Neural Systems and Rehabilitation
  Engineering}, 26\penalty0 (8):\penalty0 1481--1494, 2018.
\newblock \doi{10.1109/TNSRE.2018.2850308}.

\bibitem[Li et~al.(2020)Li, Liu, Cui, Guo, Huang, and Hu]{Li8995501}
Yang Li, Yu~Liu, Wei-Gang Cui, Yu-Zhu Guo, Hui Huang, and Zhong-Yi Hu.
\newblock Epileptic seizure detection in eeg signals using a unified
  temporal-spectral squeeze-and-excitation network.
\newblock \emph{IEEE Transactions on Neural Systems and Rehabilitation
  Engineering}, 28\penalty0 (4):\penalty0 782--794, 2020.
\newblock \doi{10.1109/TNSRE.2020.2973434}.

\bibitem[Zhao et~al.(2023)Zhao, Chu, He, Xue, Jia, Xu, and
  Zheng]{ZHAO2023104441}
Yanna Zhao, Dengyu Chu, Jiatong He, Mingrui Xue, Weikuan Jia, Fangzhou Xu, and
  Yuanjie Zheng.
\newblock Interactive local and global feature coupling for eeg-based epileptic
  seizure detection.
\newblock \emph{Biomedical Signal Processing and Control}, 81:\penalty0 104441,
  2023.
\newblock ISSN 1746-8094.
\newblock \doi{https://doi.org/10.1016/j.bspc.2022.104441}.
\newblock URL
  \url{https://www.sciencedirect.com/science/article/pii/S1746809422008953}.

\bibitem[Rukhsar et~al.(2022)Rukhsar, Tiwari, and Panda]{salim10040195}
Salim Rukhsar, Anil~Kumar Tiwari, and Samhita Panda.
\newblock Deep optimized electrodes and frequency bands in the phase space for
  identification of seizures.
\newblock In \emph{2022 IEEE 19th India Council International Conference
  (INDICON)}, pages 1--5, 2022.
\newblock \doi{10.1109/INDICON56171.2022.10040195}.

\end{thebibliography}

\end{document}